\documentclass{webofc}
\usepackage[varg]{txfonts}   
\begin{document}
\title{QCD confronts heavy-flavor and exotic hadrons}
%
%
\author{\firstname{Sasa} \lastname{Prelovsek}\inst{1,2}\fnsep\thanks{\email{sasa.prelovsek@ijs.si}} 
}

\institute{Faculty of Mathematics and Physics, University of Ljubljana, Slovenia 
\and
         Jozef Stefan Institute, Ljubljana, Slovenia 
          }

\abstract{%
   A review of QCD-based theory approaches to study the heavy-flavor and exotic hadrons is given. The focus is on the results from lattice QCD and from  lattice QCD complemented by effective field theories. Both approaches are first briefly introduced and applied in few examples. Then the status   of various hadrons sectors is presented. 
}
\maketitle
\section{Introduction}
\label{intro}

Today we know from experiment and theory that hadrons  with the following minimal quark ($q$) and gluon ($G$) contents exist:  mesons $\bar qq$, baryons $qqq$, tetraquarks $\bar qq\bar qq$, pentquarks $\bar qqqqq$ and hybrid mesons $\bar qGq$.  The first two sectors correspond to  conventional hadrons, while the last three are referred to as exotic hadrons. 

This is a brief review of how well QCD-based theory approaches confront heavy-flavor and exotic hadrons. The focus is on the results from lattice QCD and from  lattice QCD complemented by effective field theories. The two approaches are first briefly introduced and applied in a few examples for pedagogical purposes. Then  a journey over various hadron sectors is taken in order to convey the status of their understanding at present.  The temperature is zero throughout the review. 

\section{Theoretical approaches and examples}

The lattice QCD and effective field theory methods to study hadron spectroscopy have been recently reviewed in e.g. \cite{Brambilla:2019esw,Brambilla:2021mpo,Mai:2022eur}. 

\vspace{0.2cm}

  {\bf Effective field theories}
  
  Here the interactions of hadronic degrees of freedom are based on a  general Lagrangian with the symmetries of QCD. Each term comes  in general with an unknown parameter. 
 The Lagrangian is Taylor-expanded in a small quantity (eg. $1/m_b, ~p,~ E, ..$), and the  terms up to a certain order are taken into account.  

\vspace{0.2cm}

  {\bf Lattice QCD}

 The physics information on a hadron (below, near, or above threshold) is commonly extracted from the energies $E_n$ of QCD eigenstates $|n\rangle$ on a finite and discretized lattice in Eucledian space-time. The eigen-energies $E_n$ are determined from the time-dependence of  the correlation functions 
$\langle O_i (t) O_j^\dagger (0)\rangle$
 $=\sum_{n} \langle O_i|n\rangle  ~e^{-E_n t} \langle n|O^\dagger_j\rangle~,$
  where operators $O$ create/annihilate the hadron system with a given quantum number of interest.

 \underline{ Strongly stable hadrons well below threshold}: The energy of a strongly-stable hadron with zero momentum directly gives a hadron mass if this hadron is significantly below the threshold. Masses of these hadrons have already been determined  and  agree well with the experiment. For example, the precision studies of low-lying charmonia and bottomonia, that account also for QED and all systematic uncertainties, render impressive agreement   for masses and leptonic decay constants  \cite{Hatton:2021dvg,Koponen:2022hvd}. 
  
\vspace{0.2cm}

 \begin{figure}[htb]
\centering
\includegraphics[width=0.90\textwidth,clip]{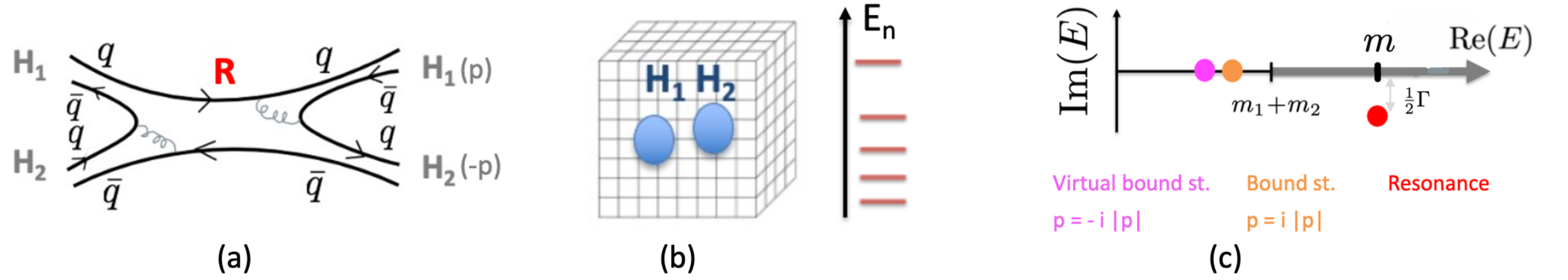}  
\caption{ Extracting   resonances and near-threshold bound states from one-channel scattering. }
\label{fig:1}
\end{figure}

\underline{  Resonances and bound states near threshold}:
In the energy region near or above threshold, the masses of bound-states and resonances have to be inferred from the scattering of two hadrons $H_1H_2$, which is encoded in the    scattering matrix $T(E)$  (Fig. \ref{fig:1}a).    The simplest example is a one-channel (elastic) scattering  in partial wave $l$, where   the scattering matrix $T(E)$ has size $1\times 1$.     L\"uscher has shown that the energy  $E$ of a two-hadron eigenstate in finite volume $L$  (Fig. \ref{fig:1}b) gives the scattering matrix $T(E)$  at that energy in infinite volume  \cite{Luscher:1991cf}.  This relation and 
 its generalizations are at the core of extracting rigorous  information about the scattering from the lattice simulations.  It leads  to $T(E)$  for real $E$  above and somewhat below $H_1H_2$ threshold, which is then analytically continued to the complex energy plane. A pole  in $T(E)$ indicates  the presence of a state, while its position renders its mass $m=\mathrm{Re}(E)$ and the width $\Gamma=-2~\mathrm{Im}(E)$, as sketched in Fig. \ref{fig:1}c . A resonance corresponds to a pole away from the real axes. A bound state corresponds to a pole below the threshold: the state is referred to the bound state if the pole occurs for positive imaginary momenta $p=i|p|$ and a virtual bound state if it occurs for $p=-i|p|$, where $p$   denotes the magnitude of the 3-momentum in the center-of-mass frame. The majority of the scattering studies are still performed at $m_\pi>m_\pi^{phy}$ and the applied values are provided in the original references.

    \begin{figure}[htb]
\centering
\includegraphics[width=0.90\textwidth,clip]{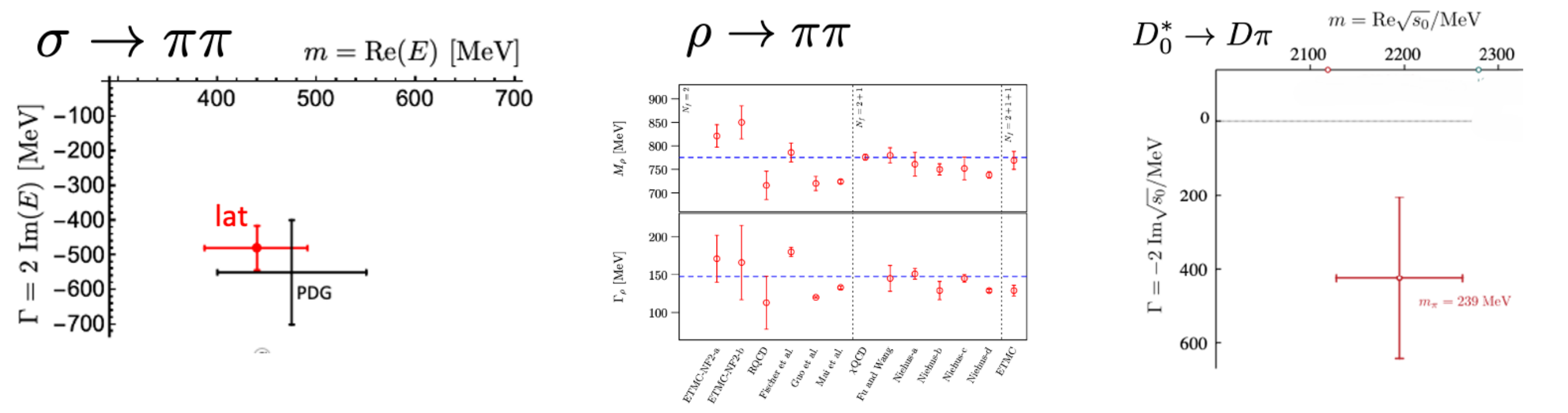}  
\caption{ Pole positions for the scalar resonances $\sigma$  \cite{Guo:2018zss} and $D_0^*$ \cite{Gayer:2021xzv}, together with  a compilation of resonances parameters for $\rho$ \cite{Mai:2022eur}. }
\label{fig:2}
\end{figure}

 Most of the resonances that decay strongly only via one decay channel have already been  extracted  from one-channel scattering, with  three examples given in Fig. \ref{fig:2}.   The $\rho$ is the only resonance that was simulated by a number of groups, and the resulting resonance parameters agree with the experiment.   The pole of $\sigma$ was extrapolated to physical $m_\pi$ using UChPT \cite{Guo:2018zss} and agrees with the pole position based on the experimental data. The 
PDG mass of the scalar $D_0^*$  is nearly degenerate with the mass of $D_{s0}^*$, which presented a puzzle, while the pole mass of $D_0^*$ from lattice simulations \cite{Gayer:2021xzv,Mohler:2012na}\footnote{The BW fit  of $D_0^*$ at real energies rendered $2.32(2)~$ GeV  \cite{Mohler:2012na}, while the corresponding pole in the complex plane based on the same lattice data is at $2.12(3)~$GeV.} is lower,   rendering  it a more natural partner of $D_{s0}^*$. 

\begin{figure}[htb]
\centering
\includegraphics[width=0.80\textwidth,clip]{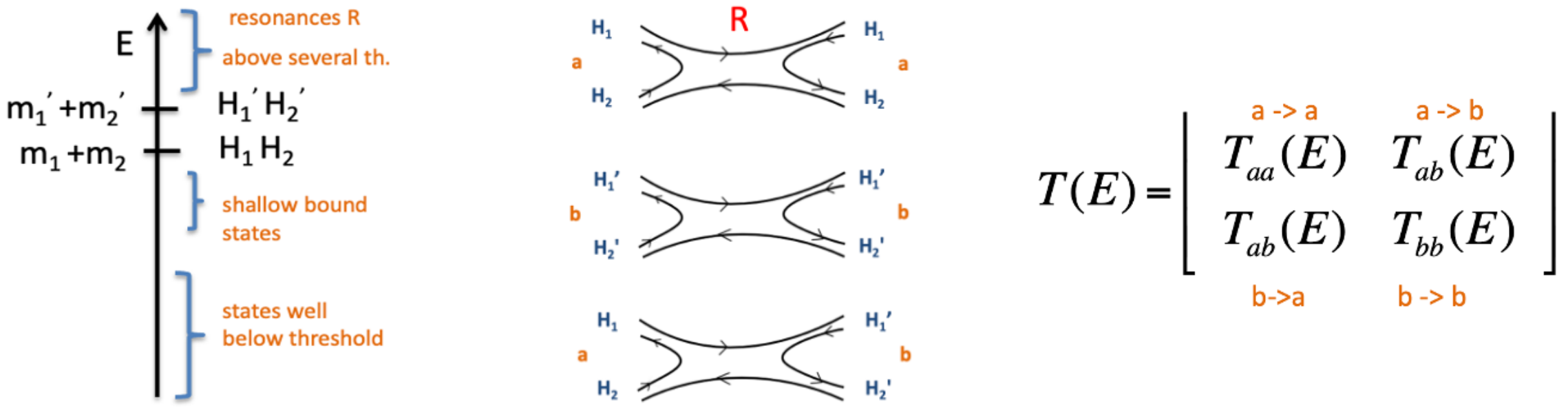}  
\caption{ A resonance that  decays via two  channels $R\to H_1H_2,~H_1^\prime H_2^\prime$   has to be inferred from the scattering of two coupled channels.    }
\label{fig:3}
\end{figure}
 
 The resonances that  decay via several strong decay channels $R\to H_1H_2,~H_1^\prime H_2^\prime, ..$ have to be extracted from the coupled-channel scattering sketched in Fig. \ref{fig:3}.  The scattering matrix for two coupled channels contains three unknown functions of energy.  It is customary to parametrize their energy dependence in order to extract them from the eigen-energies using the L\"uscher's formalism. Many resonances composed of the light quarks have already been addressed by the Hadron Spectrum Collaboration, eg \cite{Dudek:2014qha,Briceno:2017qmb}.  Concerning systems with  heavy quarks, only charmed  \cite{Moir:2016srx} and charmonium-like    \cite{Prelovsek:2020eiw} resonances were extracted, and the resulting physics conclusions   will be discussed in the next section.

\section{Various hadron sectors}

The majority of  the discovered exotic hadrons contain heavy quarks, as those are more likely to form quasi-bound states due to small kinetic energies. Most of them decay strongly, and  the theoretical challenge to study them increases with the number of decay channels.

\begin{figure}[htb]
\centering
\includegraphics[width=0.99\textwidth,clip]{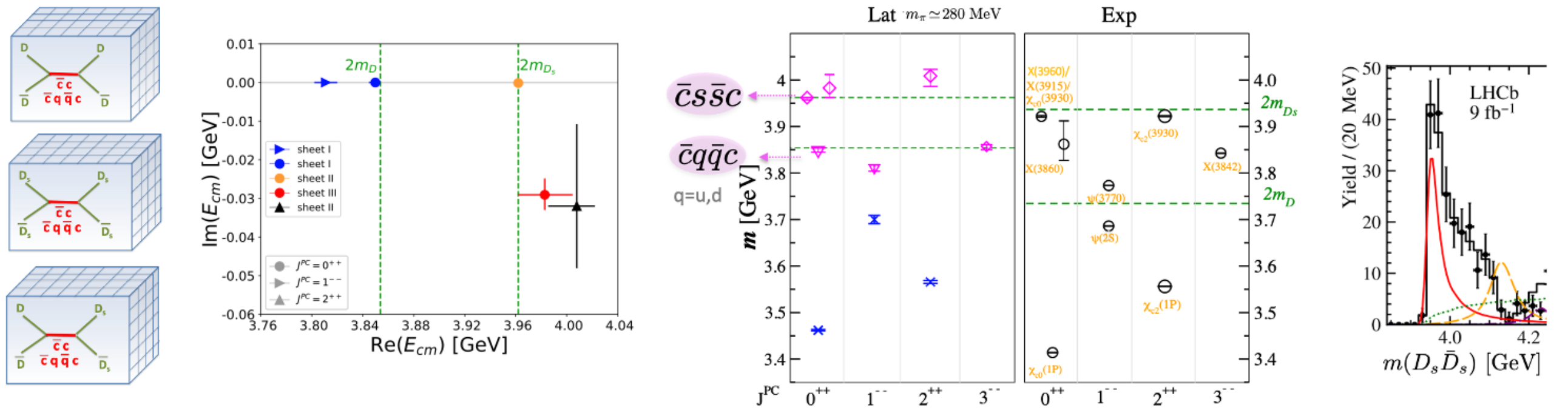}  
\caption{   Charmonium system with $I=0$: poles and masses from the lattice simulation of the scattering \cite{Prelovsek:2020eiw,Piemonte:2019cbi,Prelovsek:2021jev} and the LHCb discovery of a $X(3960)$ \cite{LHCb:2022vsv} composed of $\bar cs\bar sc$ which was predicted by the lattice simulation \cite{Prelovsek:2020eiw}.    }
\label{fig:4}
\end{figure}

\vspace{0.2cm}

{\bf Hadrons  with a heavy quark and antiquark: $\mathbf{\bar Q Q}$, $\mathbf{\bar QQ\bar qq^\prime}$}

\vspace{0.1cm}

The spectrum of charmonium-like  resonances and bound states with $I=0$ extracted from the coupled channel $\bar DD-\bar D_sD_s$ study \cite{Prelovsek:2020eiw,Piemonte:2019cbi,Prelovsek:2021jev}  is shown in Fig. \ref{fig:4}. All the states except for two (indicated by magenta arrows) appear to be conventional charmonia $\bar cc$ and the resulting masses and decay widths are in reasonable agreement with the measured ones. In addition, two exotic scalar states are predicted near both thresholds. The heavier  one has a large coupling to  $\bar D_s D_s$ and a small coupling to $\bar DD$ - it very likely corresponds to a state $X(3960)$ composed of $\bar cs\bar sc$ very recently discovered by LHCb \cite{LHCb:2022vsv}. The state near $\bar DD$ has not been claimed by experiments, while it is found in a recent dispersive reanalysis \cite{Deineka:2021aeu} of the experimental data. Both states may be dominated by the molecular Fock component and dominantly held together by the exchange of the vector mesons  $\rho,\omega,\phi$ as predicted in \cite{Gamermann:2006nm} and investigated  later in e.g. \cite{Dong:2021juy,Bayar:2022dqa}.

The charmonium-like states with explicitly exotic flavor $\bar cc\bar ud$ are challenging for a rigorous lattice study as they require a simulation of at least three coupled channels. All available lattice studies \cite{Prelovsek:2014swa,Cheung:2017tnt,CLQCD:2019npr} render eigen-energies that are close to the  non-interacting ones, which might already constrain some model interpretations, while the  solid conclusions on these interesting systems are still pending.  

\begin{figure}[htb]
\centering
\includegraphics[width=0.99\textwidth,clip]{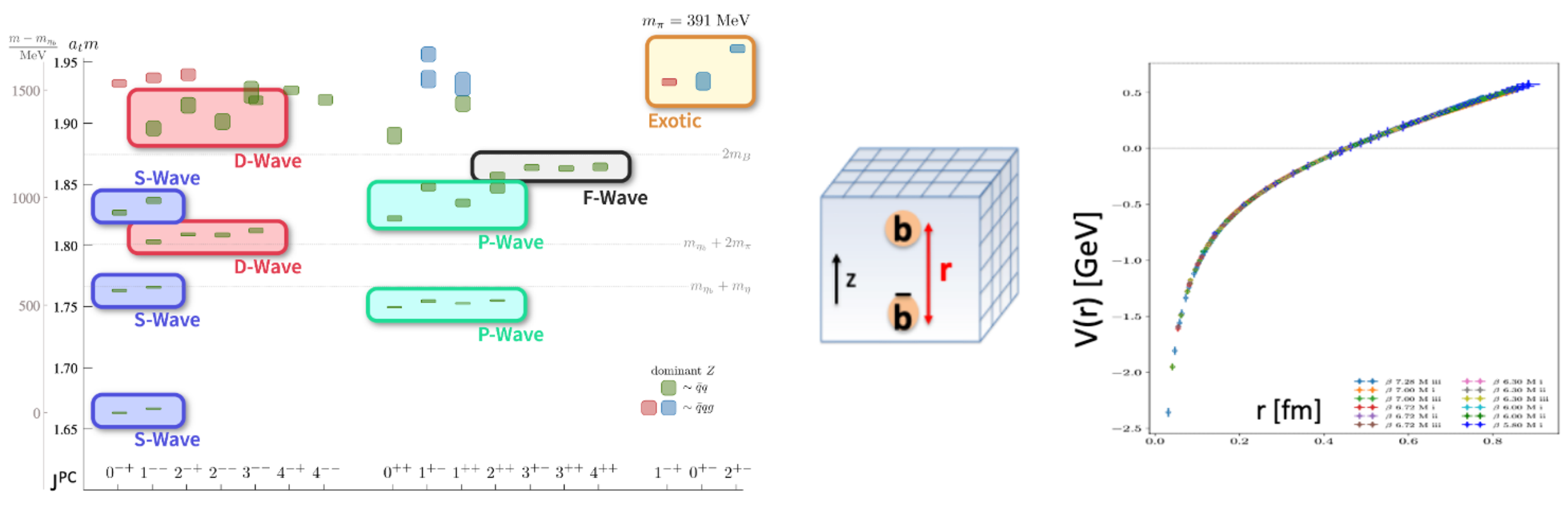}  
\caption{   The $\bar bb$ spectrum from \cite{Ryan:2020iog} and an example of $\bar bb$ static potential from \cite{Brambilla:2022het}.   }
\label{fig:5}
\end{figure}

 The conventional bottomonium spectrum $\bar bb$    \cite{Ryan:2020iog} is shown in Fig. \ref{fig:5}a, where threshold effects and strong decays  of higher-lying states are omitted. It also predicts hybrids $\bar bGb$ that await experimental discovery. 
 
 \vspace{0.1cm}
 
\begin{figure}[htb]
\centering
\includegraphics[width=0.99\textwidth,clip]{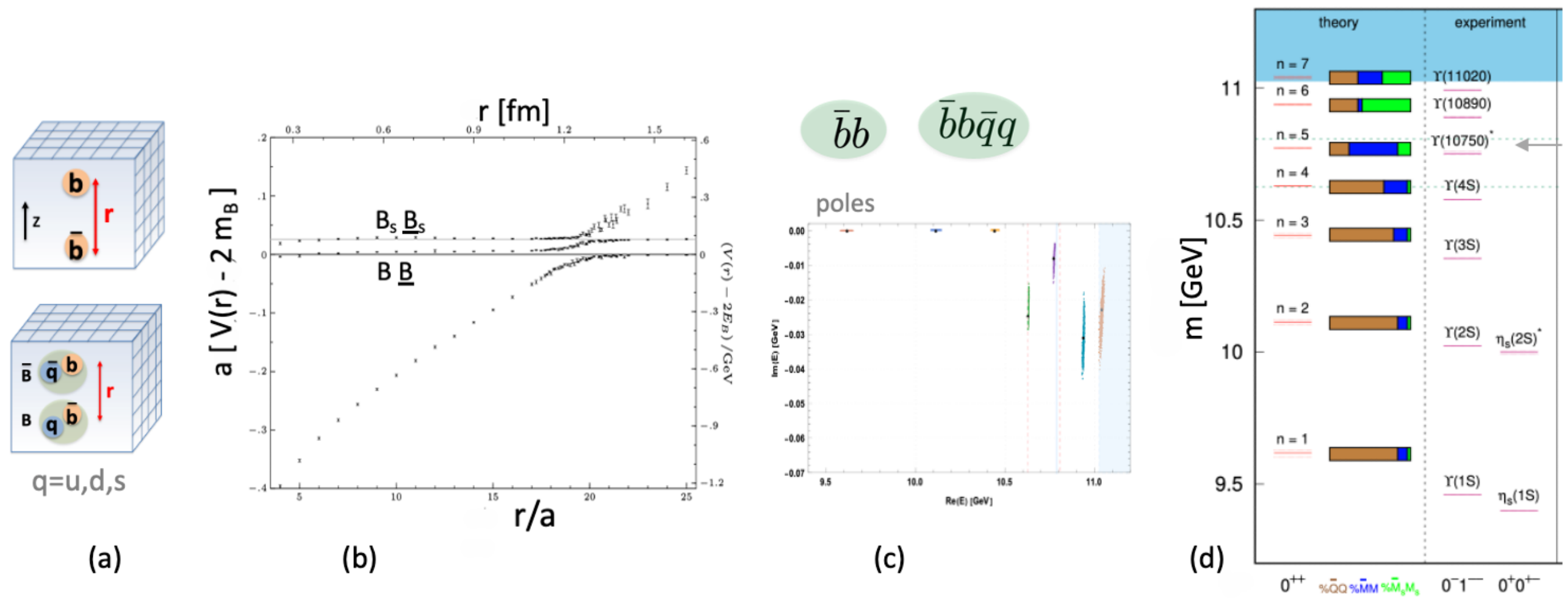}  
\caption{   (a,b) The static potential  of the quarkonium system that accounts also for the coupling to a pair of heavy mesons  \cite{Bulava:2019iut}. (c,d) Poles and composition of the bottomonium-like states \cite{Bicudo:2022ihz} from analogous static potential \cite{Bali:2005fu}.   }
\label{fig:6}
\end{figure}

\begin{figure}[htb]
\centering
\includegraphics[width=0.99\textwidth,clip]{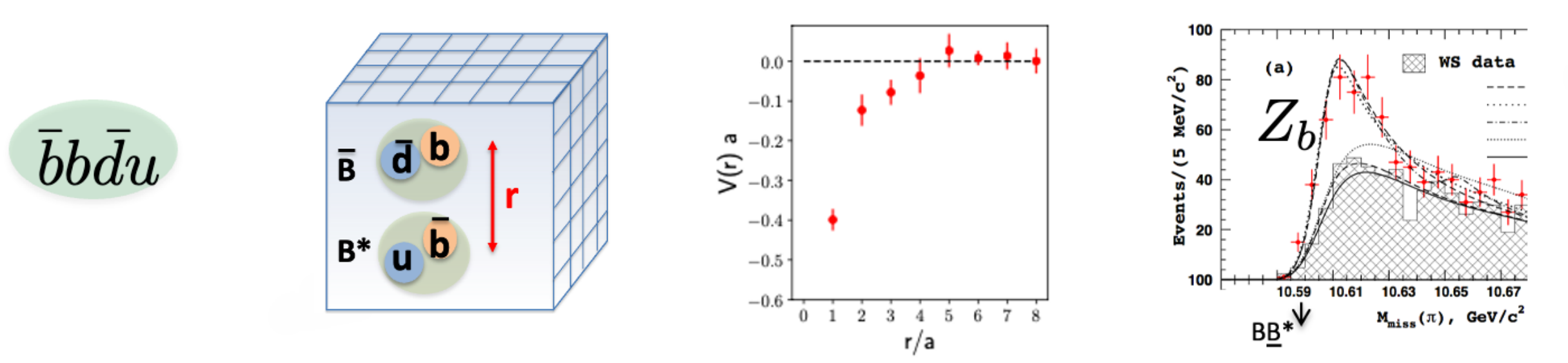}  
\caption{ The static potential between $B$ and $\bar B^*$ is attractive  \cite{Prelovsek:2019ywc}  and is likely responsible for the existence of $Z_b\simeq  \bar bb\bar du$ \cite{Belle:2015upu}.     }
\label{fig:7}
\end{figure}

\begin{figure}[htb]
\centering
\includegraphics[width=0.8\textwidth,clip]{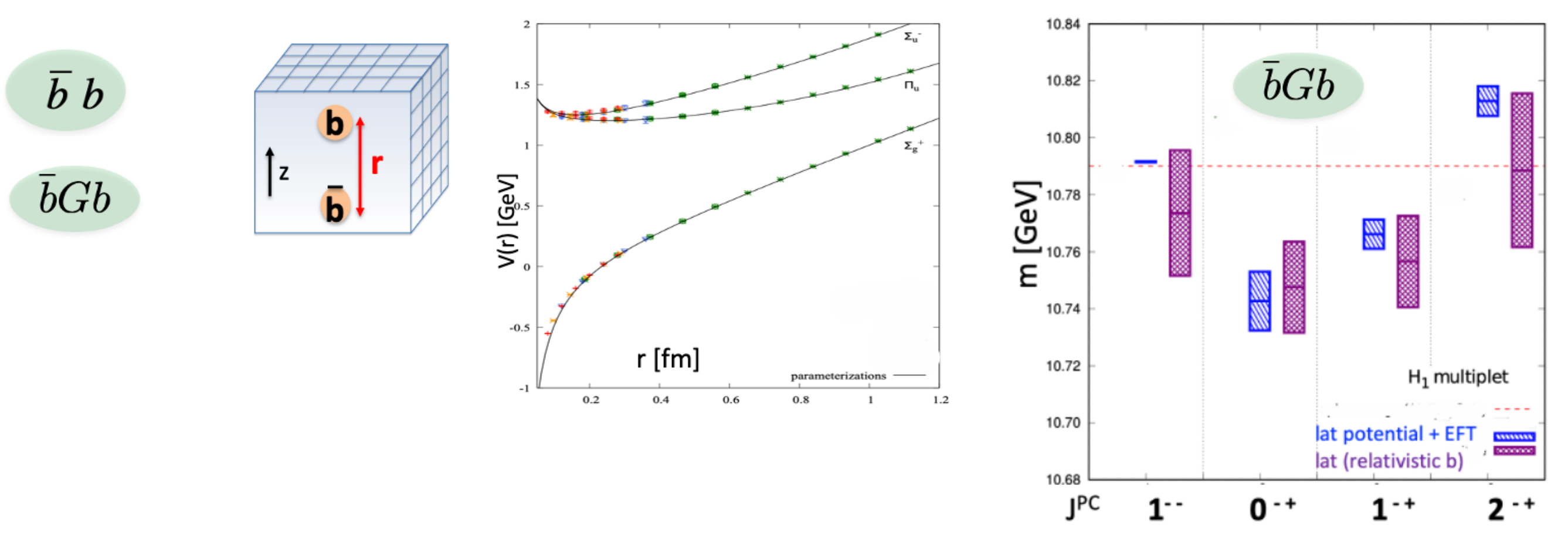}  
\caption{   Left: The   potentials related to  hybrids in quenched QCD   from a recent lattice simulation with fine lattice spacings \cite{Schlosser:2021wnr}. Right: Masses of $\bar bGb$ hybrids (update of \cite{Brambilla:2018pyn,Brambilla:2019jfi}) from older potentials  \cite{Juge:1997nc}   (blue) and from relativistic $b$ quarks \cite{Ryan:2020iog} (violet).    }
\label{fig:8}
\end{figure}

\begin{figure}[htb]
\centering
\includegraphics[width=0.5\textwidth,clip]{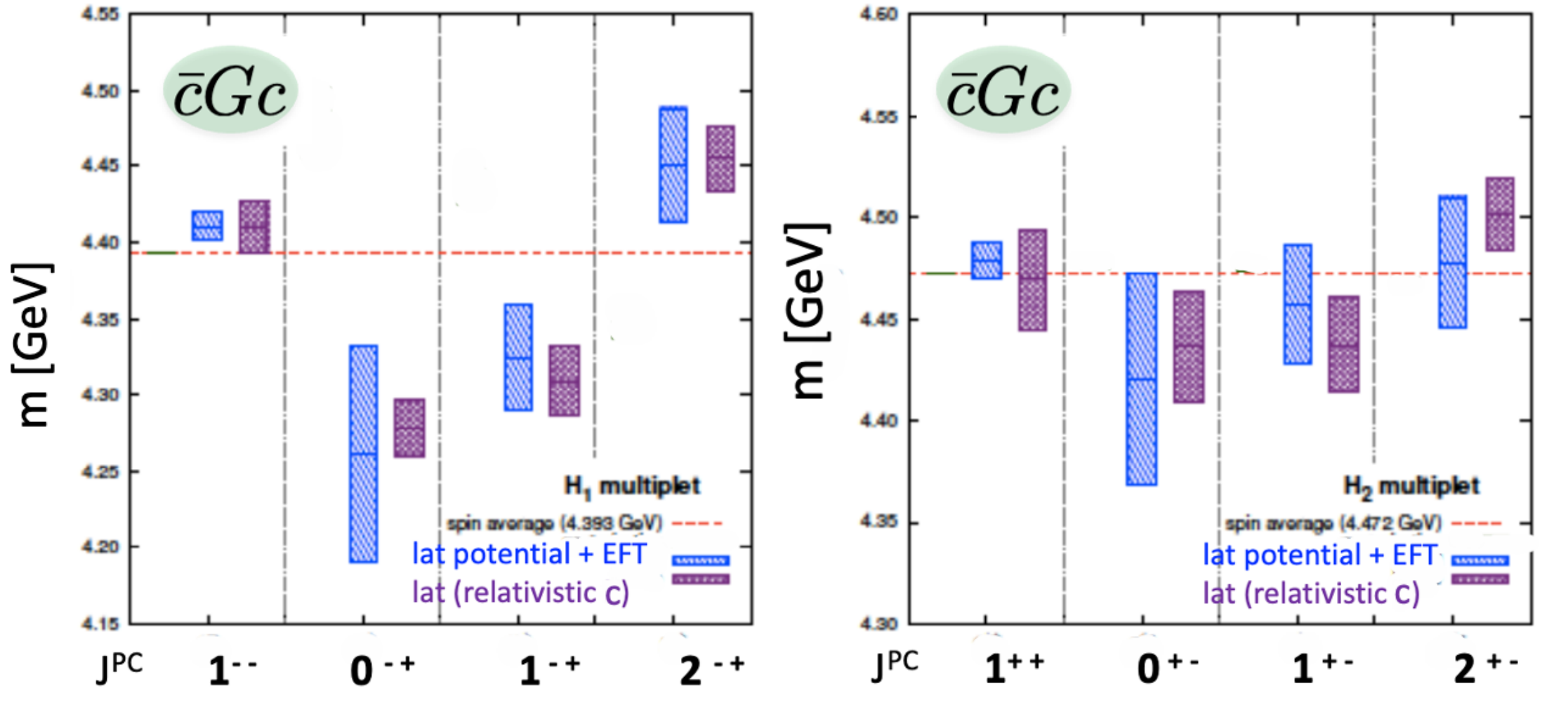}  
\caption{  Masses of $\bar cGc$ hybrids from  hybrid potentials \cite{Brambilla:2018pyn,Brambilla:2019jfi}  (blue) and from relativistic $c$ quarks \cite{Cheung:2016bym} (violet).      }
\label{fig:9}
\end{figure}

 The systems with two heavy quarks  and additional light degrees of freedom can be investigated also via the Born-Oppenheimer approach. The static energies for a fixed distance between heavy quarks render the potential $V(r)$. This is applied via the effective field theories in the Schr\"odinger equation to study the motion of the heavy quarks. The aim is to  determine whether bound states or resonances form.
 \begin{itemize}
 \item
  The $\bar bb$ static potential has been determined by many groups (Fig. \ref{fig:5}). It supports  the confinement and allows the determination of $\alpha_s$ and  the low-lying $\bar bb$ spectrum. 
  \item
   The static potential for the same $I=0$ system that accounts also for the coupling to a pair of heavy mesons \cite{Bulava:2019iut}  is shown in Fig. \ref{fig:6}b.  Coupled-channel Schr\"odinger equation with analogous  potential (but from an earlier calculation \cite{Bali:2005fu}) renders the poles related to bottomonium-like states  and their composition   in terms of $\bar bb$, $\bar BB$ and $\bar B_sB_s$  \cite{Bicudo:2022ihz} (Figs. \ref{fig:6}c,d).  These are dominated by the conventional Fock component $\bar bb$, except for state $n\!=\!5$,  which is likely related to unconventional $\Upsilon(10750)$ discovered by Belle. Another recent Born Oppenheimer study of this coupled system was performed in \cite{TarrusCastella:2022rxb}. 
   \item The observed $Z_b$ resonances with flavor content $\bar bb\bar du$ are challenging for rigorous treatment since the lowest decay channel is $\Upsilon_b \pi$, while they reside at the higher threshold $B\bar B^*$. This was taken into account in  the extraction of the potential between $B$ and $\bar B^*$ in Fig. \ref{fig:7}, which is  attractive at small distances  \cite{Peters:2016wjm,Prelovsek:2019ywc,Sadl:2021bme}. This attraction is likely responsible for the existence of the exotic $Z_b$. 
   \item The excited potentials for $\bar bb$  with certain spin-parities in Fig. \ref{fig:8} are relevant for hybrid  mesons $\bar b Gb$. The   masses from these potentials within the Born-Oppenheimer approach agree with those obtained using relativistic $b$-quarks. An analogous comparison for $\bar cGc$ is shown in Fig. \ref{fig:9}.     
   \end{itemize}

\vspace{0.2cm}

{\bf Hadrons with two heavy quarks:  $\mathbf{QQ\bar q\bar q^\prime}$  }

\vspace{0.1cm}

\begin{figure}[htb]
\centering
\includegraphics[width=0.99\textwidth,clip]{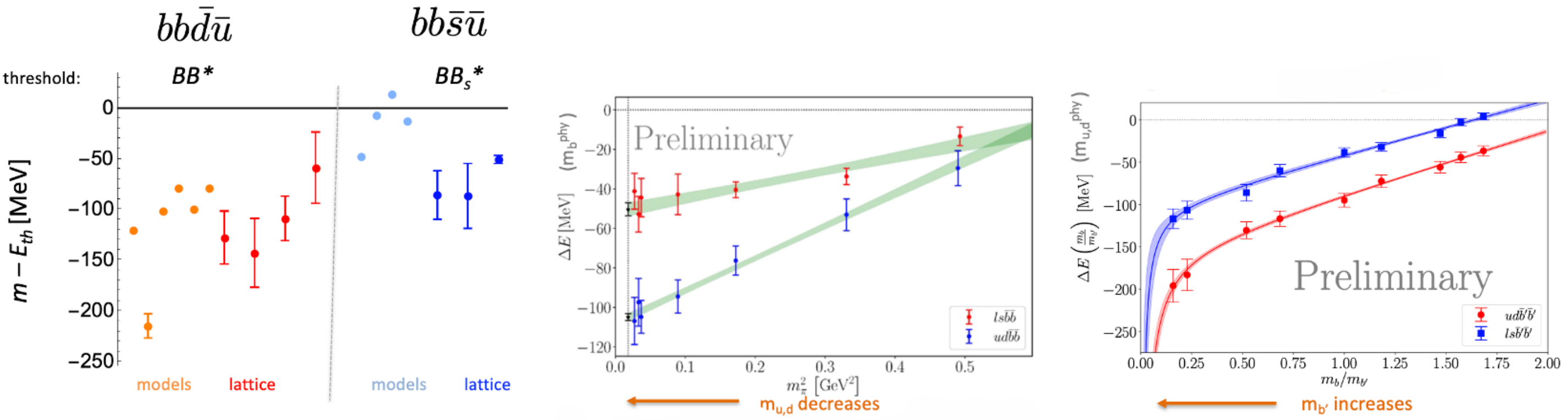}  
\caption{ Left: The binding energies of  $bb\bar u\bar d$ with $I=0$ and $J^P=1^+$ from models \cite{Eichten:2017ffp,Karliner:2017qjm,Ebert:2007rn,Janc:2004qn} and lattice \cite{Leskovec:2019ioa,Junnarkar:2018twb,Colquhoun:2022dte,Bicudo:2016ooe}. The binding energies of    $bb\bar s\bar d$  with $J^P=1^+$ from models \cite{Eichten:2017ffp,Park:2018wjk,Ebert:2007rn} and lattice \cite{Meinel:2022lzo,Junnarkar:2018twb,Colquhoun:2022dte} are also provided (cited in order from left to right).     Right:  The dependence of the binding energy on $m_b$ and $m_{u,d}$ \cite{Francis:2018jyb,Colquhoun:2022dte}.  }
\label{fig:10}
\end{figure}

 The   $bb\bar u\bar d$ and $bb\bar s\bar d$ tetraquarks with $J^P\!=\!1^+$ are expected to reside  significantly below strong decay thresholds, as shown in Fig. \ref{fig:10}. This is a reliable conclusion based on a number of lattice simulations and model-based calculations summarized in this figure. The hadron $bb\bar u\bar d$  with such a deep binding likely has a small spatial size, which   indicates the dominance of diquark antidiquark Fock component $[bb]_{\bar 3_c}^{S=1}[\bar u\bar d]_{3_c}^{S=0}$. This is also supported  by the binding energy $\Delta E$ as a function of light and heavy quark masses  in Fig. \ref{fig:10} \cite{Francis:2018jyb,Colquhoun:2022dte}: the increase of $|\Delta E|$ with  increasing $m_b$  is in line with  the Coulomb strong potential between two $b$ quarks in a diquark; the increase of  $|\Delta E|$ with decreasing $m_{u,d}$ is in line with energy dependence of a good light diquark $[\bar u\bar d]_{3_c}^{S=0}$. So far, this  the only tetraquark where lattice finds a strong support for the dominance of the diquark antidiquark Fock component.

 Tetraquarks $QQ \bar q \bar q^\prime$ ($Q\!=\!c,b~, q\!=\!u,d,s$) with other flavors and spin-parities are expected near or above strong decay thresholds $H_1H_2$. 
 Much more care is needed to establish such states  theoretically.   In the lattice formulation, for example, the lowest  eigenenergy  close to $m_{H_1}+m_{H_2}$ can indicate a presence of a bound state or a resonance, but it can arise also from two nearly free hadrons $H_1H_2$ in a finite volume.  In order to prove the existence of a state, one needs to extract the scattering matrix and establish a pole in it (Fig. \ref{fig:1}).

\begin{figure}[htb]
\centering
\includegraphics[width=0.99\textwidth,clip]{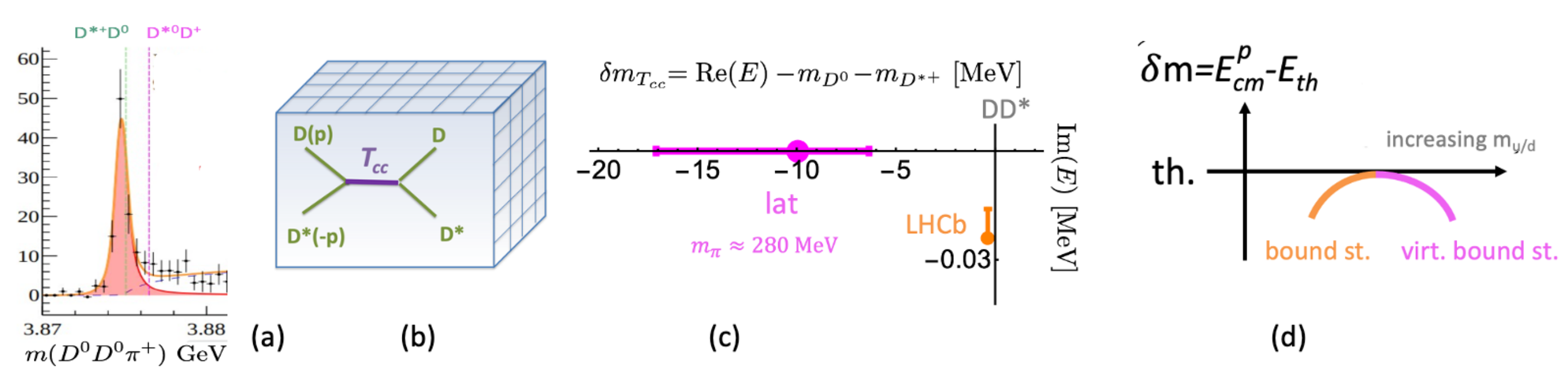}  
\caption{(a) The LHCb discovery of the doubly-charmed tetraquark $T_{cc}$ \cite{LHCb:2021vvq}; (b,c)  The lattice simulation \cite{Padmanath:2022cvl} of $DD^*$ scattering renders a virtual bound state pole below the threshold, likely related to  $T_{cc}$. (c) A sketch of $T_{cc}$ binding energy as a function of $m_{u/d}$ according to a simple toy model \cite{Padmanath:2022cvl}.  }
\label{fig:11}
\end{figure}

The doubly charm tetraquark $T_{cc}=cc\bar u\bar d$ was discovered by LHCb in 2021 just $0.4$ MeV below $DD^*$ threshold  \cite{LHCb:2021vvq}, it has   $I\!=\!0$ and most likely $J^P\!=\!1^+$. It is the longest-lived hadron with explicitly exotic quark content discovered to date. 
The first lattice QCD simulation that extracted the $DD^*$ scattering amplitude  found a virtual bound state pole about $10$ MeV below $DD^*$ threshold at $m_\pi\simeq 280~$MeV \cite{Padmanath:2022cvl}. This pole is likely related to $T_{cc}$ based on the expected dependence on $m_{u/d}$, according to which  a  quasi-bound\footnote{The $T_{cc}$ discovered by LHCb would be a bound state in the absence of decays $D^*\to D\pi$ and $T_{cc}\to DD\pi$. }  state discovered by LHCb turns to a virtual bound state at heavier $m_{u/d}$ \cite{Padmanath:2022cvl},  as sketched in Fig. \ref{fig:11}d. A subsequent lattice study of $DD^*$ scattering for $I=0$  and $I=1$ confirms  these findings and  indicates that the attraction in $I=0$ channel is dominated by the $\rho$ meson exchange  \cite{Chen:2022vpo}. Dyson-Schwinger approach also renders $T_{cc}$ below the threshold \cite{Santowsky:2021bhy}. 

\vspace{0.2cm}

{\bf Hadrons with a single heavy quark  }

\vspace{0.1cm}

\begin{figure}[htb]
\centering
\includegraphics[width=0.85\textwidth,clip]{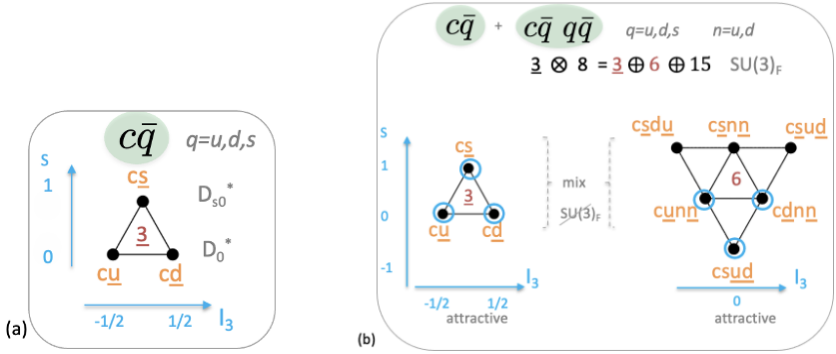}  
\caption{ The flavor contents of the scalar charmed mesons according to the quark model (a) and the new paradigm (b).  }
\label{fig:12}
\end{figure}

The charmed scalar mesons  would form a $SU(3)$ flavor triplet in Fig. \ref{fig:12}a according to the quark model. However, a new paradigm is supported by the effective field theories based on HQET and ChPT \cite{Kolomeitsev:2003ac,Du:2017zvv,Albaladejo:2016lbb,Lutz:2022enz}, combined with the lattice results on sectors  $S\!=\!1$ \cite{Mohler:2013rwa,Lang:2014yfa,Bali:2017pdv,Cheung:2020mql}, $S\!=\!0$ \cite{Mohler:2012na,Cheung:2016bym,Gayer:2021xzv} and $S\!=\!-1$ \cite{Cheung:2020mql}  as well as  experimental data \cite{Du:2017zvv,Du:2020pui}.   According to this paradigm, the spectrum features $c\bar q$ as well as $c\bar q~\bar qq$ Fock components ($q=u,d,s$). The latter decomposes to the multiplets $\bar 3\oplus 6\oplus 5$ in the $SU(3)$ flavor limit as shown in Fig.   \ref{fig:12}b.   The attractive interactions within the anti-triplet and the sextet suggest the existence of hadrons with flavors indicated by cyan circles in Fig. \ref{fig:12}b.  The breaking of $SU(3)$ mixes certain states among these representations. The states on the upper row of the sextet mix with the repulsive 15-plet. This leads to two poles for $I=1/2$ charmed mesons. The lower one resides at $2.1\!-\!2.2~$GeV in agreement with the lattice simulations \cite{,Gayer:2021xzv,Mohler:2012na} and it is a natural partner of  $D_{s0}^*(2317)$.  The heavier pole at $2.4\!-\!2.5~$GeV is suggested by the EFT re-analysis \cite{Du:2017zvv,Albaladejo:2016lbb,Lutz:2022enz}  of the lattice   \cite{Cheung:2016bym} and  experimental data. The lattice simulation of the exotic channel $D\bar K=cs\bar u\bar d$ suggests a virtual bound state below the threshold, which could be a partner of the heavier $X(2900)$ with the same flavor, that was discovered by the LHCb \cite{Cheung:2020mql}.

\vspace{0.2cm}

{\bf Di-baryons  with heavy quarks}

\vspace{0.1 cm}

\begin{figure}[htb]
\centering
\includegraphics[width=0.5\textwidth,clip]{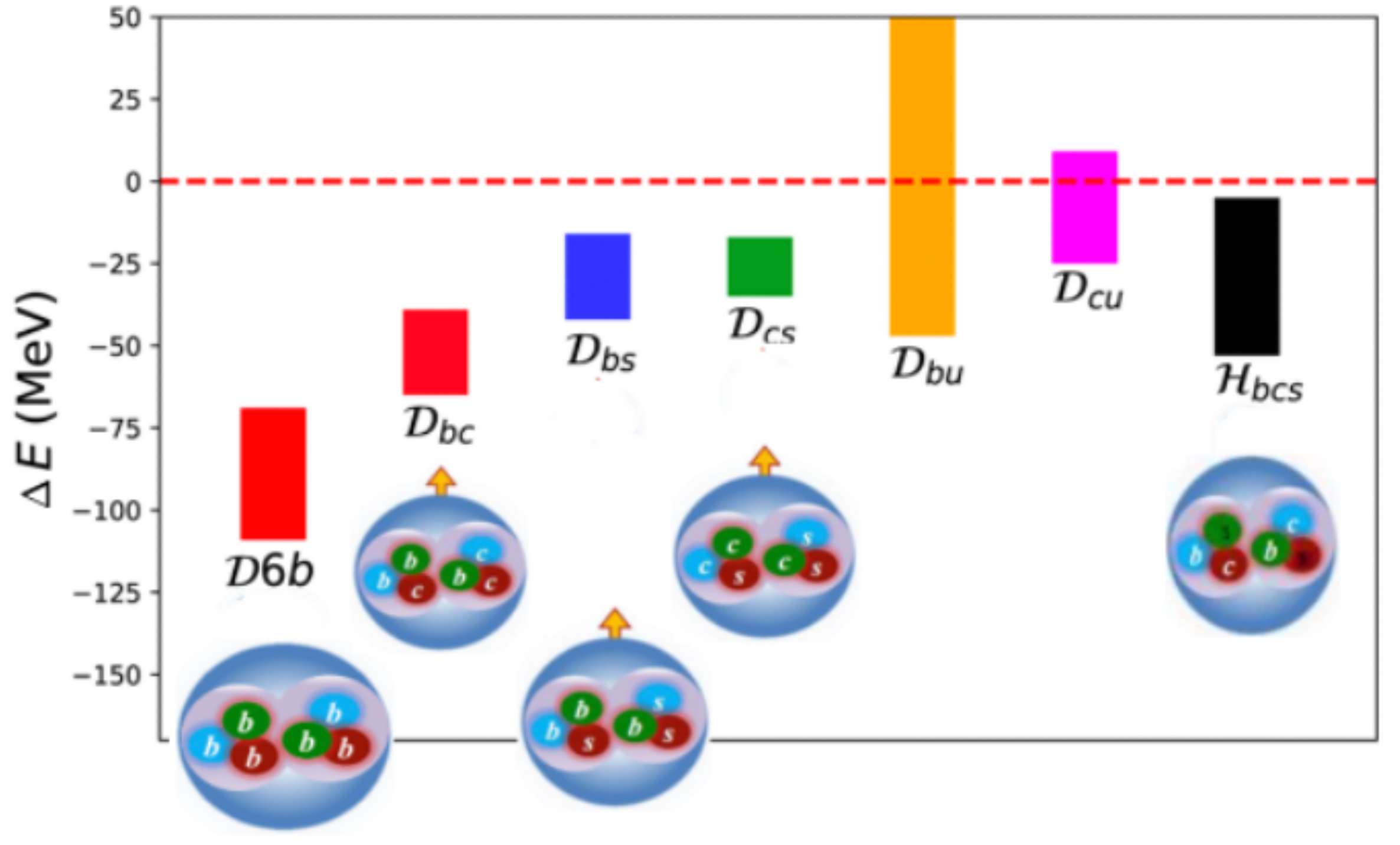}  
\caption{ Binding energies of dibaryons with heavy quarks according to the lattice simulations \cite{Junnarkar:2019equ,Mathur:2022nez,Junnarkar:2022yak}. }
\label{fig:13}
\end{figure}

The binding energies of the dibaryons that contain heavy quarks were extracted from the lattice simulations \cite{Junnarkar:2019equ,Mathur:2022nez,Junnarkar:2022yak} and are summarized in Fig.  \ref{fig:13}. The most  beautiful dibaryon is also the most strongly bound \cite{Mathur:2022nez}, while the binding decreases as some of the $b$ quarks are replaced by the lighter quarks.  A given strong potential  namely  binds more easily the heavier quarks with smaller kinetic energy. Experimental confirmation of these dibaryons represents a significant experimental challenge.

\section{Conclusions}

Experiments have provided great discoveries of new conventional as well as around thirty exotic hadrons. I have reviewed the theoretical challenge to understand the spectroscopic properties of 
various hadron sectors from first principles. I focused on the results based on lattice QCD and those where lattice QCD is supplemented by the effective field theory approaches. 

These  approaches render properties of all hadrons that are strongly stable, as well as most of the hadrons that are slightly below the strong decay threshold or decay strongly via one decay channel. Lattice simulations have rigorously treated only a few resonances that decay   via two or three decay channels and contain  heavy quarks. The theoretical challenge increases with more open decay channels. The  results suggest that certain near-threshold tetraquarks ($cc\bar u \bar d$, $D\bar D$ and $D_s\bar D_s$) might be bound via the molecular binding mechanism via the exchange of the light vector mesons, while the deeply bound $bb\bar u\bar d$ is likely due to the diquark antidiquark binding mechanism. One binding mechanism can not explain all the exotic hadrons, while  there is at least one viable mechanism for all the discovered states in the literature. 

Many of the experimentally discovered conventional and exotic hadrons, e.g. $P_c$, $Z_c(4430)$, $X(6900)$,... , are too challenging for rigorous theory treatment at present since they lie high above the lowest threshold and decay strongly via many decay channels. 

In order to gain a deeper understanding of the mechanisms responsible for the existence of the exotic hadrons, it is important to identify channels and energy regions that feature such hadrons and can be reliably investigated with both lattice QCD and experiment. A detailed study of spectroscopic properties and structure on both sides could lead to conclusions that might apply more generally. 

\vspace{0.4cm}

{\bf Acknowledgments}

I would like to thank my colleagues, in particular M. Padmanath and S. Collins, for the enjoyable and fruitful collaboration. I am grateful to colleagues in CLS for the joint effort in the generation of the gauge field ensembles, which form a basis for  our recent computations. The support from the Slovenian Research Agency ARRS (research core funding No. P1-0035) is acknowledged.


\end{document}